\documentclass[conference]{llncs}
\usepackage[]{inputenc}
\usepackage[T1]{fontenc}
\usepackage{placeins}
\usepackage{coqdoc}
\usepackage{amsmath,amssymb}
\usepackage{amssymb}
\usepackage{mathtools}
\usepackage{amsfonts}
\usepackage{algorithmic}
\usepackage{hyperref}
\usepackage{courier}
\usepackage{graphicx}
\graphicspath{{images/}}
\usepackage{float}

\begin{document}

\title{Machine-Checked Proofs For Realizability Checking Algorithms}
\author{Andreas Katis\inst{1} \and
Andrew Gacek\inst{2} \and Michael W. Whalen\inst{3}}
\institute{\inst{3}Department of
Computer Science and Engineering\\
University of Minnesota, 200 Union Street, Minneapolis, MN 55455, USA\\
\email{katis001@umn.edu},\email{whalen@cs.umn.edu}
\and Rockwell Collins Advanced Technology
Center\\
400 Collins Road NE, Cedar Rapids, IA, 52498, USA\\
\email{andrew.gacek@gmail.com}
}

\maketitle

\begin{abstract}
{\em Virtual integration} techniques focus on building architectural models of systems that can be analyzed early in the design cycle 
to try to lower cost, reduce risk, and improve quality of complex embedded systems.  Given appropriate architectural descriptions, assume/guarantee 
contracts, and compositional reasoning rules, these techniques can be used to prove important safety properties about the architecture prior to 
system construction.  For these proofs to be meaningful, each leaf-level component contract must be {\em realizable}; i.e., it is possible to 
construct a component such that for any input allowed by the contract assumptions, there is some output value that the component can produce that 
satisfies the contract guarantees.

We have recently proposed (in~\cite{Katis15:realizability}) a contract-based realizability checking algorithm for assume/guarantee
contracts over infinite theories supported by SMT solvers such as linear integer/real arithmetic and uninterpreted functions.
In that work, we used an SMT solver and an algorithm similar to k-induction to establish the realizability of a contract, and 
justified our approach via a hand proof.  Given the central importance of realizability to our virtual integration approach, we 
wanted additional confidence that our approach was sound.  This paper describes a complete formalization of the approach in the 
Coq proof and specification language.  During formalization, we found several small mistakes and missing assumptions in our reasoning. 
Although these did not compromise the correctness of the algorithm used in the checking tools, they point to the value of machine-checked 
formalization.  In addition, we believe this is the first machine-checked formalization for a realizability algorithm.

\end{abstract}

\section{Introduction}

An ongoing effort at Rockwell Collins and The University of Minnesota has explored algorithms and tools for compositional proofs of correctness.  The idea is to support hierarchical design and analysis of complex system architectures and co-evolution of requirements and architectures at multiple levels of abstraction~\cite{Whalen13:WhatHow:TwinPeaksIEEESoftware}.
We have created the AGREE reasoning framework~\cite{NFM2012:CoGaMiWhLaLu} to
support compositional assume/guarantee contract reasoning over system architectural models written in AADL.

The soundness of the compositional argument requires that each leaf-level
component contract is {\em realizable}; i.e., it is possible to construct a
component such that for any input allowed by the contract assumptions, there is
some output value that the component can produce that satisfies the contract
guarantees.  Unfortunately, without engineering support it is all too easy to
write contracts of leaf-level components that can't be realized.  
When applying our tools in both industrial and classroom settings, this issue has led to incorrect compositional ``proofs'' of systems; in fact the goal of producing a compositional proof can lead to engineers modifying component-level requirements such that they are no longer possible to implement.
In order to make our approach reasonable for practicing engineers, tool support must be provided for checking realizability.

The notion of realizability has been well-studied for many years
~\cite{Pnueli89,Bohy12,Hamza10,Chatterjee07,Gunter00,patcas2014system},
both for component synthesis and checking correctness of propositional temporal logic requirements.  Checking realizability for contracts involving theories, on the other hand, is still an open problem.
In recent work~\cite{Katis15:realizability}, we described a new approach for
checking realizability of contracts as a Satisfiability Modulo Theories (SMT)
problem and demonstrated its usefulness on several examples.  Our approach is similar to
k-induction~\cite{sheeran2000checking} over quantified formulas.  In that work, we provided
hand-proofs for several aspects of two algorithms related to the soundness of the approach with respect to both proofs and counterexamples.

Unfortunately, hand proofs of complex systems often contain errors.  Given the
criticality of realizability checking to our tool chain and the soundness of our
computational proofs, we would like a higher level of assurance than hand proofs
can provide.  In this paper, we provide a formalization of machine-checked proofs of correctness that ensure that the proposed
realizability algorithms will perform as expected, using the Coq
proof assistant.\footnote{The Coq file is available at
\url{https://github.com/andrewkatis/Coq/blob/master/realizability/Realizability.v}}
The facilities in Coq, notably mixed use of induction and co-induction, make the
construction of the proofs relatively straightforward.
This approach illustrates how interactive theorem proving and SMT solving can be used together in a profitable way.  Interactive theorem proving is used for describing the soundness of the checking algorithm (described in this paper).  The algorithm is then implemented using a SMT solver, which can automatically solve complex verification instances.

The main contribution of this paper is, therefore, the first machine-checked formalization (to our knowledge) of a realizability checking algorithm.  This is an important problem for both compositional verification involving virtual integration and component synthesis.  In addition, the formalization process exposed errors regarding our initial definitions, including necessary assumptions to one of the main theorems to be proved and an error in the definition of realizability itself.  While these errors did not ultimately impact the correctness of the algorithm, they underscore the importance of machine-checked proof.

In Section~\ref{sec:coq} we provide information on the Coq proof assistant.
Section~\ref{sec:realizability} contains the necessary informal background
towards understanding our realizability checking approach.
Sections~\ref{sec:definitions} and~\ref{sec:algorithm} describe the definitions
and theorems that were used both for defining realizability and the algorithms.
In Section~\ref{sec:implementation} we provide details on the algorithm's
implementation. Finally, in Section~\ref{sec:discussion}
we discuss our experience from the process of defining realizability and the various changes that were made along the way, and we report our conclusions in
Section~\ref{sec:conclusion}.

\section{The Coq Proof Assistant}
\label{sec:coq}

Coq\footnote{The Coq Proof
Assistant is available at \url{https://coq.inria.fr/}} is an interactive tool
used to formalize mathematical expressions and algorithms, and prove theorems
regarding their correctness and functionality~\cite{Coqmanual}. The tool was
a result of the work on the calculus of constructions~\cite{coquand1985constructions}. Its uses in the context of computer science vary, such as being a tool to represent the structure of a programming language and its characteristics, as well as to prove the correctness of underlying procedures in compilers.
Compared to other mainstream interactive theorem provers, Coq is a tool that
provides support on several aspects, such as the use of
dependent types, as opposed to the Isabelle theorem
prover~\cite{paulson1989foundation}, and proof by reflection, which is not
supported by the PVS proof assistant~\cite{cade92-pvs}.
A particularly essential feature is the
tool's support for inductive and coinductive
definitions. Definitions
using the $Inductive$ type in Coq represent a least fixpoint of the
corresponding type and are always accompanied by an induction principle, which
is implicitly used to progress through a proof by applying induction on the definition.
$CoInductive$ definitions, on the other hand, represent a greatest fixpoint to
their type. They describe a set containing every finite or infinite instance of
that type, and their proofs are essentially infinite processes, built in a
one-step fashion and requiring the existence of a guard condition that needs to
hold for them to remain well-formed. Coinductive definitions allow a natural
expression of infinite traces, which are central to our formalization of
realizability, and are tedious to prove with hand-written proofs.

\section{Realizability Checking}
\label{sec:realizability}

In~\cite{Katis15:realizability} we presented our approach to the problem of
realizability checking, introducing an algorithm involving the use of theories,
a concept that, to the best of our knowledge, has yet to be examined.
The realizability checks are defined over {\em assume-guarantee contracts}.
Informally, {\em assumptions} describe the expectations of the component on its environment, usually in terms of component inputs.  The {\em guarantees} describe the properties that will hold with respect to component outputs given that the assumptions are met.
A contract {\em holds} on an infinite trace if either the assumption is violated or the guarantee holds throughout the trace.  

To illustrate, consider a system with a single integer input $in$ and output $out$ and a contract consisting of no assumptions and two guarantees: $out = 2*in$ and $out \geq 0$.  This contract is not realizable.  At issue is the behavior of the system if $in < 0$.  In this case, the output of the system must both be positive and equal to $2*in$, which is not possible.  While this example is trivial, it can be very difficult to determine whether a contract involving dozens or hundreds of assumptions and guarantees is realizable.  In~\cite{Katis15:realizability}, we describe two large-scale compositional reasoning examples (one medical device and one flight control system) that contained unrealizable leaf-level contracts that were previously unknown that were detected by our tools.

Informally, a $realizable\ contract$ is one for which there exists a
{\em transition system} that correctly and completely implements the
contract.  By ``correctly'' we mean that the transition system always
produces outputs that satisfy the guarantees as long as the
assumptions have always been met, and by ``completely'' we mean that the
transition system never deadlocks on an input, so long as the assumptions have
always been met.  We will make these definitions precise in the next section.

This definition, while providing the proper theoretical basis for realizability,
is not actually useful for constructing our checking algorithm.  At issue is
that our current algorithm provides no way to construct this `witness'
transition system (doing so would solve the general problem of program synthesis
over contracts with theories, which we are currently researching).  We therefore
propose an alternative definition, according to which a contract is realizable if there exists a {\em viable path} consisting of {\em viable states}. A viable state is one where, for any inputs that satisfy the assumptions, there are outputs that satisfy the guarantees and lead to another viable state. This alternative definition requires that the contract be able to start in a viable state.

To derive checking algorithms from first principles, we first demonstrate that
the two definitions (transition systems and viability) are equivalent.  We can then use the viable definition as the basis of an algorithm for realizability checking.  This algorithm consists of a {\em base check}, which ensures that there exists a finitely viable state for paths of length at least $n$, and an {\em extend check} to show that all the valid paths can be further extended in response to any input. Unfortunately, the complexity of the base check does not allow for an SMT solver to handle it efficiently. Because of this, we propose a simplified version of the algorithm including a base check that ensures the extendability of every valid path consisting of viable states.  This check is only guaranteed sound with regard to 'realizable' results, that is, it may generate ``false positives'' in which the tool declares a contract unrealizable when in fact it can be realized.  In early experiments, however, the tool results have been accurate.

\section{Formalization in Coq}
\label{sec:formalization}

\begin{figure*}
\centering
\includegraphics[width=.6\textwidth]{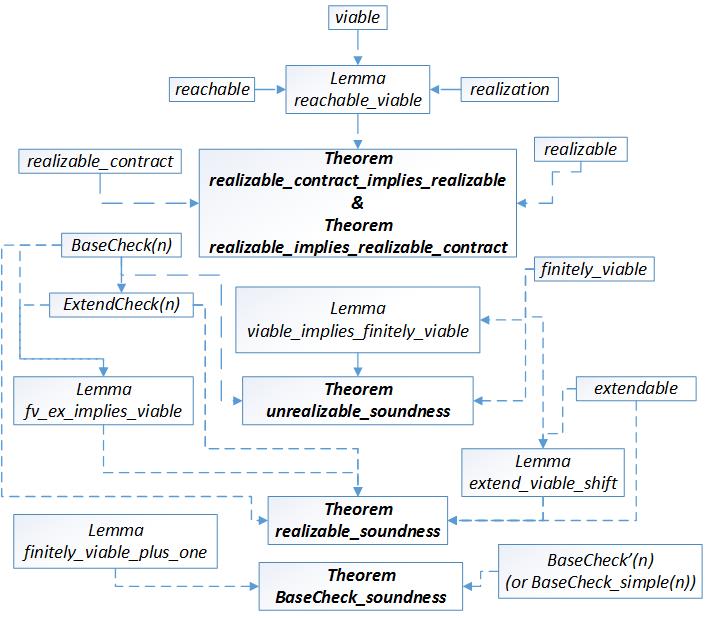}
\caption{Proof Graph}
\label{fig:pg}
\end{figure*}

In the next two subsections, we will describe the formalization and proofs of
these ideas in Coq.  Section~\ref{sec:definitions} will describe the
definitions of realizability, while Section~\ref{sec:algorithm} will
describe the algorithms for realizability checking and their proofs of adequacy
with respect to the definitions.  To provide a graphical overview of the proof
process, Figure~\ref{fig:pg} describes the connections between the various
definitions, lemmas, and theorems in our work.

\subsection{Definitions}
\label{sec:definitions}

The types $state$ and $inputs$ are used to represent a state, and
a given set of inputs.  We use Coq's $Prop$ definition to
describe the logical propositions regarding the component's $transition$
$system$ through a set $I$ of $initial$ states and the $transition$
relation $T$ between two states and a set of inputs. Finally, the contract is
defined by its $assumption$ and $guarantee$, with the latter being implicitly
referenced by a pair of initial and transitional guarantees ($iguarantee$ and $tguarantee$). The
corresponding definitions in Coq are shown below.  Note that we do not expect that a contract would be defined over all variables in the transition system -- rather its outputs -- but we do not make any distinction between internal state variables and outputs in the formalism.  This way, we can use state variables to, in some cases, simplify statements of guarantees.

\begin{coqdoccode}
\begin{itemize}
\item \coqdocnoindent
\coqdockw{Inductive} \coqdocvar{inputs} : \coqdockw{Type} :=\coqdoceol
\coqdocnoindent
\coqdocvar{input} : \coqdocvar{id} \ensuremath{\rightarrow} \coqdocvar{nat} \ensuremath{\rightarrow} \coqdocvar{inputs}.\coqdoceol
\coqdocemptyline
\item \coqdocnoindent
\coqdockw{Inductive} \coqdocvar{state} : \coqdockw{Type} :=\coqdoceol
\coqdocnoindent
\coqdocvar{st} : \coqdocvar{id} \ensuremath{\rightarrow} \coqdocvar{nat} \ensuremath{\rightarrow} \coqdocvar{state}.\coqdoceol
\coqdocemptyline
\item \coqdocnoindent
\coqdockw{Definition} \coqdocvar{initial} := \coqdocvar{state} \ensuremath{\rightarrow} \coqdockw{Prop}.\coqdoceol
\coqdocemptyline
\item \coqdocnoindent
\coqdockw{Definition} \coqdocvar{transition} := \coqdocvar{state} \ensuremath{\rightarrow} \coqdocvar{inputs} \ensuremath{\rightarrow} \coqdocvar{state} \ensuremath{\rightarrow} \coqdockw{Prop}.\coqdoceol
\coqdocemptyline
\item \coqdocnoindent
\coqdockw{Definition} \coqdocvar{iguarantee} := \coqdocvar{state} \ensuremath{\rightarrow} \coqdockw{Prop}.\coqdoceol
\coqdocemptyline
\coqdocnoindent
\item \coqdockw{Definition} \coqdocvar{tguarantee} := \coqdocvar{state}
\ensuremath{\rightarrow} \coqdocvar{inputs} \ensuremath{\rightarrow} \coqdocvar{state} \ensuremath{\rightarrow} \coqdockw{Prop}.\coqdoceol
\coqdocemptyline
\item \coqdocnoindent
\coqdockw{Definition} \coqdocvar{assumption} := \coqdocvar{state} \ensuremath{\rightarrow} \coqdocvar{inputs} \ensuremath{\rightarrow} \coqdockw{Prop}.\coqdoceol
\coqdocemptyline
\end{itemize}
\end{coqdoccode}

A state $s$ is $reachable$ with respect to the given assumptions if
there exists a path from an initial state to $s$, while each transition in the
path is satisfying the assumptions. Given a contract $(A,(G_I,G_T))$, a
transition system $(I,T)$ is its $realization$ if the following four conditions
hold:

\begin{enumerate}
\item $\forall s.~ I(s) \Rightarrow G_I(s)$
\item $\forall s, i, s'.~ reachable_A(s) \land A(s, i) \land T(s, i,
  s') \Rightarrow G_T(s, i, s')$
\item $\exists s.~ I(s)$
\item $\forall s, i.~ reachable_A(s) \land A(s, i) \Rightarrow
  \exists s'.~ T(s, i, s')$
\end{enumerate}

Finally, we define that a given contract is $realizable$, if the existence of
a transition system, which is a realization of the contract, is proved.
The formalized definitions in Coq for the $reachable$ state, the $realization$
of a contract and whether it is $realizable$ follow.

\begin{coqdoccode}
\begin{itemize}
\item \coqdocnoindent
 \coqdockw{Inductive} \coqdocvar{reachable}
(\coqdocvar{s} : \coqdocvar{state}) (\coqdocvar{I} : \coqdocvar{initial})
(\coqdocvar{T} : \coqdocvar{transition}) (\coqdocvar{A} :
\coqdocvar{assumption}) : \coqdockw{Prop} :=\coqdoceol
\coqdocnoindent \quad
\coqdocvar{rch} : 

\coqdocnoindent \quad \quad
((\coqdocvar{I} \coqdocvar{s}) \ensuremath{\lor} 

\coqdocnoindent \quad \quad
((\coqdoctac{\ensuremath{\exists}} (\coqdocvar{s'} :
\coqdocvar{state}) (\coqdocvar{inp} : \coqdocvar{inputs}), \coqdoceol
\quad \quad \quad (\coqdocvar{reachable}
\coqdocvar{s'}
\coqdocvar{I} \coqdocvar{T} \coqdocvar{A}) \ensuremath{\land} (\coqdocvar{A}
\coqdocvar{s'} \coqdocvar{inp}) \ensuremath{\land} (\coqdocvar{T}
\coqdocvar{s'} \coqdocvar{inp} \coqdocvar{s}))) \ensuremath{\rightarrow}

\coqdocnoindent \quad \quad 
\coqdocvar{reachable} \coqdocvar{s} \coqdocvar{I} \coqdocvar{T}
\coqdocvar{A}).\coqdoceol
\coqdocemptyline
\item \coqdocnoindent
\coqdockw{Inductive} \coqdocvar{realization} (\coqdocvar{I} :
\coqdocvar{initial}) (\coqdocvar{T} : \coqdocvar{transition}) (\coqdocvar{A} : \coqdocvar{assumption}) (\coqdocvar{$G_I$} : \coqdocvar{iguarantee}) (\coqdocvar{$G_T$} : \coqdocvar{tguarantee}) : \coqdockw{Prop} :=\coqdoceol
\coqdocnoindent
\coqdocvar{real} : ((\coqdockw{\ensuremath{\forall}} (\coqdocvar{s} :
\coqdocvar{state}), (\coqdocvar{I} \coqdocvar{s}) \ensuremath{\rightarrow}
(\coqdocvar{$G_I$} \coqdocvar{s})) \ensuremath{\land} \coqdoceol
\quad \quad  \quad (\coqdockw{\ensuremath{\forall}} (\coqdocvar{s}
\coqdocvar{s'} :
\coqdocvar{state}) (\coqdocvar{inp} : \coqdocvar{inputs}), \coqdoceol
\quad \quad \quad ((\coqdocvar{reachable} \coqdocvar{s} \coqdocvar{I}
\coqdocvar{T} \coqdocvar{A}) \ensuremath{\land} (\coqdocvar{A} \coqdocvar{s} \coqdocvar{inp})
 \ensuremath{\land} (\coqdocvar{T}
 \coqdocvar{s}
 \coqdocvar{inp}
 \coqdocvar{s'})) \ensuremath{\rightarrow} \coqdocvar{$G_T$} \coqdocvar{s} \coqdocvar{inp} \coqdocvar{s'}) \ensuremath{\land}\coqdoceol 
 \quad \quad \quad (\coqdoctac{\ensuremath{\exists}} (\coqdocvar{s} :
 \coqdocvar{state}),
 \coqdocvar{I} \coqdocvar{s}) \ensuremath{\land} \coqdoceol
 \quad \quad \quad (\coqdockw{\ensuremath{\forall}} (\coqdocvar{s} :
 \coqdocvar{state}) (\coqdocvar{inp} : \coqdocvar{inputs}), (\coqdocvar{reachable} \coqdocvar{s} \coqdocvar{I} \coqdocvar{T} \coqdocvar{A} \ensuremath{\land} (\coqdocvar{A} \coqdocvar{s} \coqdocvar{inp})) \ensuremath{\rightarrow}

\quad \quad \quad \quad (\coqdoctac{\ensuremath{\exists}} (\coqdocvar{s'} :
\coqdocvar{state}), \coqdocvar{T} \coqdocvar{s} \coqdocvar{inp} \coqdocvar{s'}))) \ensuremath{\rightarrow}\coqdoceol

\quad \quad \coqdocvar{realization} \coqdocvar{I} \coqdocvar{T} \coqdocvar{A} \coqdocvar{$G_I$} \coqdocvar{$G_T$}.\coqdoceol
\coqdocemptyline
\item \coqdocnoindent
\coqdockw{Inductive} \coqdocvar{realizable\_contract}
(\coqdocvar{A} : \coqdocvar{assumption}) (\coqdocvar{$G_I$} :
\coqdocvar{iguarantee}) (\coqdocvar{$G_T$} : \coqdocvar{tguarantee}) :
\coqdockw{Prop} :=\coqdoceol
\coqdocnoindent
\coqdocvar{rc} : (\coqdoctac{\ensuremath{\exists}} (\coqdocvar{I} :
\coqdocvar{initial}) (\coqdocvar{T} : \coqdocvar{transition}),
\coqdocvar{realization} \coqdocvar{I} \coqdocvar{T} \coqdocvar{A}
\coqdocvar{$G_I$} \coqdocvar{$G_T$}) \ensuremath{\rightarrow} \coqdoceol
\quad \quad \coqdocvar{realizable\_contract} \coqdocvar{A} \coqdocvar{$G_I$}
\coqdocvar{$G_T$}.\coqdoceol
\coqdocemptyline
\end{itemize}
\end{coqdoccode}

While the definitions of $realization$
and $realizable\_contract$ are quite straightforward, they cannot be used
directly to construct an actual realizability checking algorithm.
Therefore, we proposed the notion of a state being $viable$ with respect to a contract, meaning that the transition system continues to be a realization of the
contract, while we are at such a state. In other words, a state is $viable$
($viable(s)$) if the transitional guarantee $G_T$ infinitely holds, given valid
inputs.
Using the definition of $viable$, a contract is $realizable$ if and only if $\exists s.~ G_I(s) \land viable(s)$.

\begin{coqdoccode}
\begin{itemize}
\item \coqdocnoindent
\coqdockw{CoInductive} \coqdocvar{viable} (\coqdocvar{s} : \coqdocvar{state})
(\coqdocvar{A} : \coqdocvar{assumption}) (\coqdocvar{$G_I$}
: \coqdocvar{iguarantee}) (\coqdocvar{$G_T$}: \coqdocvar{tguarantee}) :
\coqdockw{Prop} :=\coqdoceol
\coqdocnoindent
\coqdocvar{vbl} : (\coqdockw{\ensuremath{\forall}} (\coqdocvar{inp} :
\coqdocvar{inputs}), (\coqdocvar{A} \coqdocvar{s} \coqdocvar{inp})
\ensuremath{\rightarrow} 

\quad \quad \quad (\coqdoctac{\ensuremath{\exists}} (\coqdocvar{s'} :
\coqdocvar{state}), \coqdocvar{$G_T$} \coqdocvar{s} \coqdocvar{inp}
\coqdocvar{s'} \ensuremath{\land} \coqdocvar{viable} \coqdocvar{s'}
\coqdocvar{A} \coqdocvar{$G_I$} \coqdocvar{$G_T$})) \ensuremath{\rightarrow} \coqdoceol 

\quad \quad \coqdocvar{viable} \coqdocvar{s} \coqdocvar{A} \coqdocvar{$G_I$}
\coqdocvar{$G_T$}.\coqdoceol
\coqdocemptyline
\item\coqdocnoindent
 \coqdockw{Inductive} \coqdocvar{realizable} (\coqdocvar{A}
: \coqdocvar{assumption}) (\coqdocvar{$G_I$} : \coqdocvar{iguarantee})
(\coqdocvar{$G_T$} : \coqdocvar{tguarantee}) : \coqdockw{Prop} :=\coqdoceol
\quad \coqdocvar{rl} : (\coqdoctac{\ensuremath{\exists}} (\coqdocvar{s}
: \coqdocvar{state}), \coqdocvar{$G_I$} \coqdocvar{s} \ensuremath{\land}
\coqdocvar{viable} \coqdocvar{s} \coqdocvar{A} \coqdocvar{$G_I$} \coqdocvar{$G_T$})
\ensuremath{\rightarrow} \coqdocvar{realizable} \coqdocvar{A} \coqdocvar{$G_I$} \coqdocvar{$G_T$}.\coqdoceol
\coqdocemptyline
\end{itemize}
\end{coqdoccode}

Having a more useful definition for realizability, we need to prove the
equivalence between the definitions of $realizable\_contract$ and $realizable$. The Coq definition of
the theorem was split into two separate theorems, each for one of the two directions of the proof.
Towards the two proofs, the auxiliary lemma that, given a realization, $\forall
s.~ reachable_A(s) \Rightarrow viable(s)$ is necessary.

\begin{coqdoccode}
\begin{itemize}
  \item \coqdocnoindent
\coqdockw{Lemma} \coqdocvar{reachable\_viable} : \coqdockw{\ensuremath{\forall}}
(\coqdocvar{s} : \coqdocvar{state}) (\coqdocvar{I} : \coqdocvar{initial})
(\coqdocvar{T} : \coqdocvar{transition}) (\coqdocvar{A} :
\coqdocvar{assumption}) (\coqdocvar{$G_I$} : \coqdocvar{iguarantee})
(\coqdocvar{$G_T$} : \coqdocvar{tguarantee}),\coqdoceol
\coqdocnoindent \quad
\coqdocvar{realization} \coqdocvar{I} \coqdocvar{T} \coqdocvar{A} \coqdocvar{$G_I$} \coqdocvar{$G_T$} \ensuremath{\rightarrow} \coqdocvar{reachable} \coqdocvar{s} \coqdocvar{I} \coqdocvar{T} \coqdocvar{A} \ensuremath{\rightarrow} \coqdocvar{viable} \coqdocvar{s} \coqdocvar{A} \coqdocvar{$G_I$} \coqdocvar{$G_T$}.\coqdoceol
 \coqdocemptyline
\end{itemize}
\end{coqdoccode}

The informal proof of the lemma relies initially on the unrolling of the
$viable$ definition, for a specific state $s$. Thus, we are left to prove that
there exists another state $s'$ that we can traverse into, in addition to being
viable.
The former can be proved directly from the conditions 2 and 4 of the definition of
$realization$. For the latter, by the definition of $viable$ on $s'$ we
need to show that $s'$ is reachable. Given the definition of $reachable$ though,
we just need to prove that there exists another reachable state from which we can
reach $s'$, in one step. But we already know that $s$ is such a state, and thus
the lemma holds.

\begin{coqdoccode}
\begin{itemize}
  \item \coqdocnoindent
\coqdockw{Theorem} \coqdocvar{realizable\_contract\_implies\_realizable}
(\coqdocvar{I} :
\coqdocvar{initial}) (\coqdocvar{T} : \coqdocvar{transition}) :
\coqdockw{\ensuremath{\forall}} (\coqdocvar{A} : \coqdocvar{assumption})
(\coqdocvar{$G_I$} : \coqdocvar{iguarantee}) (\coqdocvar{$G_T$} :
\coqdocvar{tguarantee}),\coqdoceol
\coqdocnoindent \quad
\coqdocvar{realizable\_contract} \coqdocvar{A} \coqdocvar{$G_I$} \coqdocvar{$G_T$} \ensuremath{\rightarrow} \coqdocvar{realizable} \coqdocvar{A} \coqdocvar{$G_I$} \coqdocvar{$G_T$}.\coqdoceol
\coqdocemptyline
\item \coqdocnoindent
\coqdockw{Theorem} \coqdocvar{realizable\_implies\_realizable\_contract}
(\coqdocvar{I} :
\coqdocvar{initial}) (\coqdocvar{T} : \coqdocvar{transition}) :
\coqdockw{\ensuremath{\forall}} (\coqdocvar{A} : \coqdocvar{assumption})
(\coqdocvar{$G_I$} : \coqdocvar{iguarantee}) (\coqdocvar{$G_T$} :
\coqdocvar{tguarantee}),\coqdoceol
\coqdocnoindent \quad
\coqdocvar{realizable} \coqdocvar{A} \coqdocvar{$G_I$} \coqdocvar{$G_T$} \ensuremath{\rightarrow} \coqdocvar{realizable\_contract} \coqdocvar{A} \coqdocvar{$G_I$} \coqdocvar{$G_T$}.\coqdoceol
\coqdocemptyline
\end{itemize}
\end{coqdoccode}

The first part of the theorem requires us to prove that there exists a viable
state $s$ for which the initial guarantee holds. Considering that we
have a contract that is realizable under the $realizable\_contract$ definition,
we have a transition system that is a realization of the contract, and thus from
the third condition of the $realization$ definition, there exists an initial
state $s'$ for which, using the first condition, the initial guarantee
holds. Thus, we are left to prove that $s'$ is viable. But, by proving that $s'$
is reachable, we can use the $reachable\_viable$ lemma to show that $s'$ is
indeed viable.

The second direction requires a bit more effort. Assuming that we have a viable
state $s_0$ with $G_I(s_0)$ being true, we define $I(s) = (s =
s_0)$ and $T(s, inp, s') = G_T(s, inp, s') \land viable(s')$. Initially, we need
to prove the $reachable\_viable$ lemma in this context, with the additional
assumption that another viable state already exists ($s_0$ in this case).
Having done so, we need to prove that there exists a transition system that is a realization of the given
contract. Given the transition system that we defined earlier, we need to show
that each of the four conditions hold. Since $I(s) = (s =
s_0)$ and $G_I(s_0)$ hold, the proof for the first condition is trivial. Using
the assumption that $T(s, inp, s') = G_T(s, inp, s') \land viable(s')$, we can
also trivially prove the second condition, while the third condition is simply proved
by reflexivity on the state $s_0$. Finally, for the fourth condition we need to
prove that $\forall s, inp.~ reachable_A(s) \land A(s, inp) \Rightarrow
\exists s'.~ G_T(s, inp, s') \land viable(s')$. By applying the
$reachable\_viable$ lemma on the reachable state $s$ in the assumptions,
we show that $s$ is also viable, if $s_0$ is viable, which is what we assumed in
the first place. Thus, coming back into what we need to prove, and unrolling the
definition of $viable$ on $s$, we have that $\forall inp.~ A(s, inp) \Rightarrow
\exists s'.~ G_T(s, inp, s') \land viable(s')$ which completes the proof.

\subsection{Algorithms}
\label{sec:algorithm}

In this section we provide a description of the formalization and
proof of soundness of our realizability checking algorithms.
Initially, we define an under-approximation of the definition of viability, for
the finite case. Thus, a state is $finitely\_viable$ for $n$ steps
($viable_n(s)$), if the transitional guarantee $G_T$ holds for at least $n$
steps, given valid inputs.	

\begin{coqdoccode}
\begin{itemize}
  \item \coqdocnoindent
\coqdockw{Inductive} \coqdocvar{finitely\_viable} : \coqdocvar{nat}
\ensuremath{\rightarrow} \coqdocvar{state} \ensuremath{\rightarrow}
\coqdocvar{assumption} \ensuremath{\rightarrow} \coqdocvar{tguarantee}
\ensuremath{\rightarrow} \coqdockw{Prop} := \ \coqdoceol
\coqdocnoindent \quad 
\ensuremath{|}~\coqdocvar{fvnil} : \coqdockw{\ensuremath{\forall}} \coqdocvar{s}
\coqdocvar{A} \coqdocvar{$G_T$}, \coqdocvar{finitely\_viable} \coqdocvar{O}
\coqdocvar{s} \coqdocvar{A} \coqdocvar{$G_T$}\coqdoceol 
 
\coqdocnoindent \quad
\ensuremath{|}~\coqdocvar{fv} : \coqdockw{\ensuremath{\forall}} \coqdocvar{n}
\coqdocvar{s} \coqdocvar{A} \coqdocvar{$G_T$}, \coqdocvar{finitely\_viable}
\coqdocvar{n} \coqdocvar{s} \coqdocvar{A} \coqdocvar{$G_T$}
\ensuremath{\rightarrow} 

\coqdocnoindent \quad \quad \quad \quad
(\coqdockw{\ensuremath{\forall}} (\coqdocvar{inp} :
\coqdocvar{inputs}), \coqdocvar{A} \coqdocvar{s} \coqdocvar{inp}
\ensuremath{\rightarrow} 
(\coqdoctac{\ensuremath{\exists}} \coqdocvar{s'},
\coqdocvar{$G_T$} \coqdocvar{s} \coqdocvar{inp} \coqdocvar{s'}))
\ensuremath{\rightarrow} 

\coqdocnoindent \quad \quad \quad \quad
\coqdocvar{finitely\_viable} (\coqdocvar{S}
\coqdocvar{n}) \coqdocvar{s} \coqdocvar{A} \coqdocvar{$G_T$}.\coqdoceol
\coqdocemptyline
\end{itemize}
\end{coqdoccode}

In addition to the $finitely\_viable$ definition, an under-approximation of
viability is also used, called one-step extension. Therefore, a valid path
leading to a state $s$ is $extendable$ after $n$ steps, if any path from $s$, of
length at least $n$, can be further extended given a valid input.

\begin{coqdoccode}
\begin{itemize}
  \item \coqdocnoindent
\coqdockw{Inductive} \coqdocvar{extendable} : \coqdocvar{nat}
\ensuremath{\rightarrow} \coqdocvar{state} \ensuremath{\rightarrow}
\coqdocvar{assumption} \ensuremath{\rightarrow} \coqdocvar{tguarantee}
\ensuremath{\rightarrow} \coqdockw{Prop} := 

\coqdocnoindent
\ensuremath{|}~\coqdocvar{exnil} : \coqdockw{\ensuremath{\forall}}
(\coqdocvar{s} : \coqdocvar{state}) (\coqdocvar{A} : \coqdocvar{assumption})
(\coqdocvar{$G_T$} : \coqdocvar{tguarantee}), \coqdoceol
\quad \quad \quad \quad (\coqdockw{\ensuremath{\forall}} (\coqdocvar{inp} :
\coqdocvar{inputs}),
\coqdocvar{A} \coqdocvar{s} \coqdocvar{inp} \ensuremath{\rightarrow}
\coqdoctac{\ensuremath{\exists}} (\coqdocvar{s'} : \coqdocvar{state}),
\coqdocvar{$G_T$} \coqdocvar{s} \coqdocvar{inp} \coqdocvar{s'})
\ensuremath{\rightarrow} \coqdoceol
\quad \quad \quad \quad \coqdocvar{extendable} \coqdocvar{O} \coqdocvar{s}
\coqdocvar{A} \coqdocvar{$G_T$}\coqdoceol

\coqdocnoindent
\ensuremath{|}~\coqdocvar{ex} : \coqdockw{\ensuremath{\forall}} \coqdocvar{n}
\coqdocvar{s} \coqdocvar{A} \coqdocvar{$G_T$}, \coqdoceol
\quad \quad \quad (\coqdockw{\ensuremath{\forall}}
\coqdocvar{inp} \coqdocvar{s'}, \coqdocvar{A} \coqdocvar{s} \coqdocvar{inp}
\ensuremath{\land} \coqdocvar{$G_T$} \coqdocvar{s} \coqdocvar{inp}
\coqdocvar{s'} \ensuremath{\land} \coqdocvar{extendable}
\coqdocvar{n} \coqdocvar{s'} \coqdocvar{A} \coqdocvar{$G_T$})
\ensuremath{\rightarrow} \coqdoceol 
\quad \quad \quad \coqdocvar{extendable} (\coqdocvar{S}
\coqdocvar{n}) \coqdocvar{s} \coqdocvar{A} \coqdocvar{$G_T$}.\coqdoceol
\coqdocemptyline
\end{itemize}
\end{coqdoccode}

\subsubsection{An Exact Algorithm for Realizability Checking}
\label{sec:exactalgorithm}

The algorithm that we propose for realizability checking consists of
two checks.
The $BaseCheck(n)$ procedure ensures that $\exists s.~ G_I(s) \land viable_n(s)$, while $ExtendCheck(n)$ makes sure
that the given state from $BaseCheck$ is extendable for any $n$.

\begin{coqdoccode}
\begin{itemize}
  \item \coqdocnoindent
\coqdockw{Definition} \coqdocvar{BaseCheck} (\coqdocvar{n} : \coqdocvar{nat})
(\coqdocvar{A} : \coqdocvar{assumption}) (\coqdocvar{$G_I$} :
\coqdocvar{iguarantee}) (\coqdocvar{$G_T$} : \coqdocvar{tguarantee})
:=\coqdoceol
\coqdocnoindent
\coqdoctac{\ensuremath{\exists}} (\coqdocvar{s} : \coqdocvar{state}),
(\coqdocvar{$G_I$} \coqdocvar{s} \ensuremath{\land} \coqdocvar{finitely\_viable}
\coqdocvar{n} \coqdocvar{s} \coqdocvar{A} \coqdocvar{$G_T$}).\coqdoceol
\coqdocemptyline
\coqdocnoindent
\coqdockw{Definition} \coqdocvar{ExtendCheck} (\coqdocvar{n} : \coqdocvar{nat})
(\coqdocvar{A} :
\coqdocvar{assumption}) (\coqdocvar{$G_T$} : \coqdocvar{tguarantee}) :=\coqdoceol
\coqdocnoindent
\coqdockw{\ensuremath{\forall}} \coqdocvar{s} \coqdocvar{A} \coqdocvar{$G_T$}, \coqdocvar{extendable} \coqdocvar{n} \coqdocvar{s} \coqdocvar{A} \coqdocvar{$G_T$}.\coqdoceol
\coqdocemptyline
\end{itemize}
\end{coqdoccode}

Using the $BaseCheck(n)$ and $ExtendCheck(n)$, the algorithm determines the
realizability of the given contract, using the following procedure.

\begin{algorithmic}
  \FOR{$n=0$ to $\infty$}
    \IF{\NOT $BaseCheck(n)$}
      \RETURN{``unrealizable''}
    \ELSIF{$ExtendCheck(n)$}
      \RETURN{``realizable''}
    \ENDIF
  \ENDFOR
\end{algorithmic}

\hspace{+1cm}

Using the definitions of $BaseCheck$ and $ExtendCheck$, we proved the
algorithm's soundness, both for the 'unrealizable' and 'realizable' case. The
main idea behind the proof of soundness for the 'unrealizable' result is to prove the contrapositive, that is, given a realizable contract, there exists
a natural number $x$ for which $BaseCheck(x)$ holds. Unfolding
the definition of $BaseCheck(x)$, we need to show that $\exists s.~ G_I(s) \land
viable_x(s)$. Knowing that our assumption $realizable\_contract$ $A$ $G_I$
$G_T$ is equivalent to the $realizable$ definition, provides us with a state
$s'$, for which $G_I(s') \land viable(s')$ holds. Here, we need an additional lemma, according to which $\forall s, n.~
viable(s) \Rightarrow viable_n(s)$ (stated as
$viable\_implies\_finitely\_viable$ below). Thus, using the lemma on
$viable(s')$ with $n = x$, we get that $viable_x(s')$, thus completing the
proof.

\begin{coqdoccode}
\begin{itemize}
\item \coqdocnoindent
\coqdockw{Lemma} \coqdocvar{viable\_implies\_finitely\_viable} :
\coqdockw{\ensuremath{\forall}} \coqdocvar{s} \coqdocvar{A} \coqdocvar{$G_I$}
\coqdocvar{$G_T$} \coqdocvar{n},\coqdoceol
\coqdocnoindent
\coqdocvar{viable} \coqdocvar{s} \coqdocvar{A} \coqdocvar{$G_I$}
\coqdocvar{$G_T$} \ensuremath{\rightarrow} \coqdocvar{finitely\_viable}
\coqdocvar{n} \coqdocvar{s} \coqdocvar{A} \coqdocvar{$G_T$}.\coqdoceol
\coqdocemptyline
\item \coqdocnoindent
\coqdockw{Theorem} \coqdocvar{unrealizable\_soundness} :
\coqdockw{\ensuremath{\forall}}
(\coqdocvar{I} : \coqdocvar{initial}) (\coqdocvar{T} : \coqdocvar{transition})
(\coqdocvar{A} : \coqdocvar{assumption}) (\coqdocvar{$G_I$} :
\coqdocvar{iguarantee}) (\coqdocvar{$G_T$} : \coqdocvar{tguarantee}),\coqdoceol
\coqdocnoindent
(\coqdoctac{\ensuremath{\exists}} \coqdocvar{n}, \ensuremath{\lnot}\coqdocvar{BaseCheck} \coqdocvar{n} \coqdocvar{A} \coqdocvar{$G_I$} \coqdocvar{$G_T$}) \ensuremath{\rightarrow} \ensuremath{\lnot} \coqdocvar{realizable\_contract} \coqdocvar{A} \coqdocvar{$G_I$} \coqdocvar{$G_T$}.\coqdoceol
\coqdocemptyline
\end{itemize}
\end{coqdoccode}

For the soundness of the 'realizable' result, we first need to prove two
lemmas. Initially, $extend\_viable\_shift$, shows the way that
$Extend_n(s)$ can be used to shift $viable_n(s)$ forward. The proof for this lemma is done by
using induction on $n$. The base case is proved trivially, by unfolding the
definitions of $extendable$ and $finitely\_viable$ in the assumptions. For the
inductive case, we assume that the same state $s$ is extendable and finitely
viable for paths of length $n+1$, and try to prove that there exists a finitely
viable state $s'$ for paths of length $n+1$, to which we can traverse from $s$,
with the contract guarantees still holding after the transition. By
considering that $s$ is extendable for paths of length $n+1$, we can use it as
that potentially existing state in the proof, requiring that we can transition
from $s$ to itself, with the transitional guarantees staying true, and $s$
being finitely viable for paths of length $n+1$. The former is true through the
definition of $extendable$, while the second is an already given assumption by
the inductive step.

\begin{coqdoccode}
\begin{itemize}
  \item \coqdocnoindent
\coqdockw{Lemma} \coqdocvar{extend\_viable\_shift} :
\coqdockw{\ensuremath{\forall}} (\coqdocvar{s} : \coqdocvar{state})
(\coqdocvar{n} : \coqdocvar{nat}) (\coqdocvar{inp} : \coqdocvar{inputs})
(\coqdocvar{A} : \coqdocvar{assumption}) (\coqdocvar{$G_I$} :
\coqdocvar{iguarantee}) (\coqdocvar{$G_T$} : \coqdocvar{tguarantee}),\coqdoceol
\coqdocnoindent \quad
(\coqdocvar{extendable} \coqdocvar{n} \coqdocvar{s} \coqdocvar{A}
\coqdocvar{$G_T$} \ensuremath{\land} \coqdocvar{finitely\_viable} \coqdocvar{n}
\coqdocvar{s} \coqdocvar{A} \coqdocvar{$G_T$} \ensuremath{\land} \coqdocvar{A}
\coqdocvar{s} \coqdocvar{inp}) \ensuremath{\rightarrow}

\quad \quad
(\coqdoctac{\ensuremath{\exists}} \coqdocvar{s'}, \coqdocvar{$G_T$}
\coqdocvar{s} \coqdocvar{inp} \coqdocvar{s'} \ensuremath{\land}
\coqdocvar{finitely\_viable} \coqdocvar{n} \coqdocvar{s'} \coqdocvar{A}
\coqdocvar{$G_T$}).\coqdoceol
\coqdocemptyline
\item \coqdocnoindent
\coqdockw{Lemma} \coqdocvar{fv\_ex\_implies\_viable} :
\coqdockw{\ensuremath{\forall}} (\coqdocvar{s} : \coqdocvar{state})
(\coqdocvar{n} : \coqdocvar{nat}) (\coqdocvar{A} : \coqdocvar{assumption})
\coqdocvar{$G_I$} \coqdocvar{$G_T$},\coqdoceol
\coqdocnoindent
(\coqdocvar{finitely\_viable} \coqdocvar{n} \coqdocvar{s} \coqdocvar{A}
\coqdocvar{$G_T$} \ensuremath{\land} \coqdocvar{ExtendCheck} \coqdocvar{n} \coqdocvar{A} \coqdocvar{$G_T$}) \ensuremath{\rightarrow} \coqdocvar{viable} \coqdocvar{s} \coqdocvar{A} \coqdocvar{$G_I$} \coqdocvar{$G_T$}.\coqdoceol
\coqdocemptyline
\item \coqdocnoindent
\coqdockw{Theorem} \coqdocvar{realizable\_soundness} :
\coqdockw{\ensuremath{\forall}}
(\coqdocvar{I} : \coqdocvar{initial}) (\coqdocvar{T} : \coqdocvar{transition})
\coqdocvar{A} \coqdocvar{$G_I$} \coqdocvar{$G_T$},\coqdoceol
\coqdocnoindent
(\coqdoctac{\ensuremath{\exists}} \coqdocvar{n}, (\coqdocvar{BaseCheck}
\coqdocvar{n} \coqdocvar{A} \coqdocvar{$G_I$} \coqdocvar{$G_T$}
\ensuremath{\land} \coqdocvar{ExtendCheck} \coqdocvar{n} \coqdocvar{A}
\coqdocvar{$G_T$})) \ensuremath{\rightarrow}
\coqdoceol \coqdocvar{realizable\_contract} \coqdocvar{A} \coqdocvar{$G_I$}
\coqdocvar{$G_T$}.\coqdoceol
\coqdocemptyline
\end{itemize}
\end{coqdoccode}

To prove the theorem, we try to prove the equivalent for the
$realizable$ definition instead. The existence of a state for which the initial
guarantees hold is derived from the assumption that $BaseCheck$ holds for a
finitely viable state, while the proof that the same state is also viable comes
from the use of the $fv\_ex\_implies\_viable$ lemma, which is proved through the
use of $extend\_viable\_shift$.

\subsubsection{An Approximate Algorithm for Realizability Checking}
\label{sec:approxalgorithm}

Following the definition of our approach, we
noticed the problematic nature of $BaseCheck(n)$ having $2n$ quantifier
alternations, which cannot be handled efficiently by an SMT solver. To that end,
we proposed a simplified version of the $BaseCheck(n)$ procedure, called $BaseCheck'(n)$,
stated as $BaseCheck\_simple$ below.

\begin{coqdoccode}
\begin{itemize}
\item \coqdocnoindent
\coqdockw{Definition} \coqdocvar{BaseCheck\_simple} (\coqdocvar{n} :
\coqdocvar{nat}) (\coqdocvar{A} : \coqdocvar{assumption}) (\coqdocvar{$G_I$} : \coqdocvar{iguarantee}) (\coqdocvar{$G_T$} : \coqdocvar{tguarantee}) :=
\coqdocnoindent
\coqdockw{\ensuremath{\forall}} \coqdocvar{s}, (\coqdocvar{$G_I$} \coqdocvar{s})
\ensuremath{\rightarrow} \coqdocvar{extendable} \coqdocvar{n} \coqdocvar{s} \coqdocvar{A} \coqdocvar{$G_T$}.\coqdoceol
\coqdocemptyline
\item \coqdocnoindent
\coqdockw{Lemma} \coqdocvar{finitely\_viable\_plus\_one} :
\coqdockw{\ensuremath{\forall}} \coqdocvar{s} \coqdocvar{n} \coqdocvar{A}
(\coqdocvar{gi} : \coqdocvar{iguarantee}) (\coqdocvar{$G_T$} :
\coqdocvar{tguarantee}) (\coqdocvar{inp} :
\coqdocvar{inputs}),\coqdoceol
\coqdocnoindent
(\coqdocvar{extendable} \coqdocvar{n} \coqdocvar{s} \coqdocvar{A}
\coqdocvar{$G_T$} \ensuremath{\land} \coqdocvar{finitely\_viable} \coqdocvar{n}
\coqdocvar{s} \coqdocvar{A} \coqdocvar{$G_T$}) \ensuremath{\rightarrow}
\coqdoceol
\coqdocvar{finitely\_viable} (\coqdocvar{S} \coqdocvar{n}) \coqdocvar{s}
\coqdocvar{A} \coqdocvar{$G_T$}.\coqdoceol
\coqdocemptyline \coqdoceol  
\item \coqdocnoindent 
\coqdockw{Theorem} \coqdocvar{BaseCheck\_soundness} :
\coqdockw{\ensuremath{\forall}} \coqdocvar{n} \coqdocvar{A} (\coqdocvar{$G_I$} :
\coqdocvar{iguarantee}) (\coqdocvar{$G_T$} : \coqdocvar{tguarantee})
(\coqdocvar{i} :
\coqdocvar{inputs}),\coqdoceol
\coqdocnoindent
((\coqdoctac{\ensuremath{\exists}} \coqdocvar{s}, \coqdocvar{$G_I$} \coqdocvar{s}) \ensuremath{\land} 
(\coqdockw{\ensuremath{\forall}} \coqdocvar{k},
(\coqdocvar{k}\ensuremath{\le}\coqdocvar{n}) \ensuremath{\rightarrow}
\coqdocvar{BaseCheck\_simple} \coqdocvar{k} \coqdocvar{A} \coqdocvar{$G_I$}
\coqdocvar{$G_T$})) \ensuremath{\rightarrow} \coqdoceol \coqdocvar{BaseCheck}
\coqdocvar{n} \coqdocvar{A} \coqdocvar{$G_I$} \coqdocvar{$G_T$}.\coqdoceol
\end{itemize}
\end{coqdoccode}

The simplified $BaseCheck'(n)$, while being an easier instance for an
SMT solver, is not sound for the 'unrealizable' case, falsely reporting
some realizable contracts to not be so. Nevertheless, we proved the modified
algorithm's soundness for the 'realizable' result, with the use of an
auxiliary lemma.

The lemma, $finitely\_viable\_plus\_one$ simply refers to the fact that an
extendable and finitely viable state $s$, for a given number of steps $n$,
is also finitely viable for $n+1$ steps. The proof is done by induction on
$n$. The base case is trivially proved, by the definition of $finitely\_viable$,
and the assumption that $s$ is extendable. For the
inductive case, we use the inductive hypothesis, which leaves us to prove
the assumptions on a specific state $s$. The extendability is trivially shown
since we already know that $s$ is extendable for paths of length $n+1$, with
the same idea being applied to prove that $s$ is finitely viable for $n$.

Finally, the proof of soundness for the 'realizable' result of the
$BaseCheck'(n)$ procedure is done by using induction on $n$. The base case is
trivially true, using the fact that all paths of zero length are finitely
viable. The inductive step then requires us to prove that $BaseCheck(n+1)$ holds.
In order to do so, we need to construct the inductive hypothesis' assumption,
as a separate assumption to the theorem's scope.
By applying the inductive hypothesis to the newly created assumption, we have
that $BaseCheck(n)$ holds. By unrolling the definition of $BaseCheck(n)$ and
applying the lemma $finitely\_viable\_plus\_one$ on the extracted state, say
$x$, we finally prove that $x$ is extendable through the definition of
$BaseCheck'(n)$, completing the proof at the same time.

Figure~\ref{fig:pg} provides a simplified proof graph of all the necessary
definitions and partially, for graph simplicity purposes, the way that they are
used towards proving the lemmas and theorems stated in this paper.

\section{Implementation}
\label{sec:implementation}

The algorithm is now an optional feature, namely JRealizability in
JKind~\cite{jkind}, a Java implementation of the KIND 2 model checker,\footnote{You can download the KIND
model checker at \url{http://kind2-mc.github.io/kind2/}} and supports models
expressed using the Lustre language~\cite{halbwachs1991synchronous}, which are
a result of AGREE's translation process of contracts written in AADL.
A typical process for checking models in the above environment starts from
providing the corresponding Lustre program to JKind, which JRealizability uses to find a
number $n$, $n \geq 0$, such that both $BaseCheck'(n)$ and $ExtendCheck(n)$
hold.
Specifically, the model's variables and contract are being translated in the
SMT-LIB2 format, followed by the construction of each check's corresponding
query for the current value of $n$, in its negated form. The resulting SMT-LIB2
file is provided as input to the Z3 SMT solver~\cite{DeMoura08:z3}, which
attempts to answer the given query. In the case that the negated formula is unsatisfiable,
JRealizability returns a 'realizable' result. On the other hand, a satisfiable
query implies that the model is unrealizable. Consequently, the tool requests a
model, i.e. an instance of the contract's variables that reflects Z3's
result, and proceeds to construct a counterexample that describes the exact
cause of the contract's unrealizability. Finally, in those cases where the
quantified query is too difficult for Z3 to solve, an 'unknown' result is reported, both by Z3 and JRealizability.

The implementation was used in~\cite{Katis15:realizability} to verify the
correctness of contracts in terms of realizability in three different case
studies. The performance was very good for the concrete results, with the tool
exceeding its predefined timeout value for the 'unknown' ones. False positive
results ($BaseCheck'(n)$) were not found during this process, as every
unrealizable contract was manually proved to be a result of conflicts in the
provided assumptions or guarantees. A final remark is the fact that the most
critical case studies already had an implementation that was supposed to work
correctly. As such, the discovery of unrealizable contracts in these systems
eventually required a total revision of the formalized requirements defined for
each system, thus hindering the development process.

\section{Discussion}
\label{sec:discussion}

While our work on realizability is based on simple definitions, formalizing
them and refining the algorithms in Coq was non-trivial.
Proving the lemmas and theorems using Coq helped us discover minor errors in 
our informal statements. For
example, our proof of the one-way soundness theorem for the simplified $BaseCheck$ in~\cite{Katis15:realizability} lacks the necessary assumption that there exists
a state for which the initial guarantees hold. Another example is that we forgot
to include initial states in our definition of reachable states in the informal
proof.  The use of a mechanized theorem prover exposed some missing knowledge in 
the informal text, and helped us provide a more precise version of the theorem.  
Although these errors in the hand proofs did not lead to problems with our 
implementation, Coq improved both our theorems and proofs, and provided a very 
high level of assurance that our algorithm is correct.

\section{Conclusion}
\label{sec:conclusion}

The work in this paper was particularly important towards verifying our
approach and learning more about the actual functionality of the algorithm.
Interactive theorem provers like Coq provide the necessary support to define
the notions and assertions while being able to effectively prove theorems in a far more convenient and reassuring way,
in contrast to hand-written, informal proofs, especially when it comes down to
tracking formulas containing alternating quantifiers.
Furthermore, the procedure of proving the theorems in an interactive way with a tool allowed us to refine our
definitions. Additionally, the time that
was required was minimal when compared to the process of considering the
informal proofs and writing down our requirements in English.
The most important outcome was the proof of correctness of our approach that
enabled us to provide a complementary  set of definitions and proofs,
easily processed by an experienced Coq user.

To conclude, there is substantial additional work that could be performed in
 terms of fleshing out the formalisms used in the proofs for our particular
 implementation.  For example, we could define the structure and types of inputs and outputs, and describe how transition systems are realized in the AGREE tool suite.  However, the work that has been performed shows the soundness of the proof system and our algorithms with respect to proofs of realizability, allowing us to proceed with very high confidence as to the correctness of our approach.

\hspace{+1cm}

\vspace{-0.1in}
\subsubsection{Acknowledgments.}
This work was funded by DARPA and AFRL under contract 4504789784 (Secure Mathematically-Assured Composition of Control Models), and by NASA under contract NNA13AA21C (Compositional Verification of Flight Critical Systems), and by NSF under grant CNS-1035715 (Assuring the safety, security, and reliability of medical device cyber physical systems).
\vspace{-0.1in}
\bibliographystyle{IEEEtran}
\bibliography{document}

\end{document}